\documentclass[traditabstract,longauth]{aa}
\usepackage{epsfig}
\usepackage{txfonts}
\usepackage{graphicx}

\def\ms{\,m\,s$^{-1}$}         
\def\kms{\,km\,s$^{-1}$}         
\def\vsini{$v$\,sin\,$i$}      
\def\ms{\hbox{\,m\,s$^{-1}$}}         
\def\m2s2{\hbox{\,m$^{2}$\,s$^{-2}$}} 
\def\kms{\hbox{\,km\,s$^{-1}$}}       
\def\vsini{\hbox{$v$\,sin\,$i$}}      
\def\Msun{\hbox{$\mathrm{M}_{\odot}$}}             
\def\Rsun{\hbox{$\mathrm{R}_{\odot}$}}
\def\Mjup{\hbox{$\mathrm{M}_{\rm Jup}$}}
\def\Rjup{\hbox{$\mathrm{R}_{\rm Jup}$}}

\def\teff{T$_{\rm eff}$}
\def\logg{log {\it g}}

\def\mr{$M_\star^{1/3}/R_\star$}
\def\modif{}

\begin{document}
\title{Transiting exoplanets from the $CoRoT$ space mission\thanks{The $CoRoT$ space mission, 
launched on December 27th 2006, has been developed and is operated by CNES, with the 
contribution of Austria, Belgium, Brazil , ESA (RSSD and Science Programme), Germany and 
Spain. Observations made with HARPS spectrograph at ESO La Silla Observatory (184.C-0639).}}
\subtitle{XV. CoRoT-15b: a brown dwarf transiting companion}

\author{
Bouchy, F.\inst{1,2} 
\and Deleuil, M.\inst{3} 
\and Guillot, T.\inst{4} 
\and Aigrain, S.\inst{5} 
\and Carone, L.\inst{6} 
\and Cochran, W.D. \inst{7}
\and Almenara, J.M. \inst{8,9} 
\and Alonso, R.\inst{10} 
\and Auvergne, M.\inst{11} 
\and Baglin, A.\inst{11}  
\and Barge, P.\inst{3} 
\and Bonomo, A. S.\inst{3}
\and Bord\'e, P.\inst{12} 
\and Csizmadia, Sz.\inst{13} 
\and De Bondt, K.\inst{3}
\and Deeg, H.J.\inst{8,9} 
\and D\'iaz, R.F. \inst{1}
\and Dvorak, R.\inst{14}
\and Endl, M. \inst{7}
\and Erikson, A.\inst{13}
\and Ferraz-Mello, S.\inst{15}
\and Fridlund, M.\inst{16}
\and Gandolfi, D.\inst{16,17}
\and Gazzano, J.C. \inst{3}
\and Gibson, N.\inst{5}
\and Gillon, M.\inst{18} 
\and Guenther, E.\inst{17}
\and Hatzes, A.\inst{17}
\and Havel, M. \inst{4}
\and H\'ebrard, G.\inst{1} 
\and Jorda, L.\inst{3} 
\and L\'eger, A.\inst{12} 
\and Lovis, C. \inst{10}
\and Llebaria, A.\inst{3}
\and Lammer, H.\inst{19} 
\and MacQueen, P.J.\inst{7}
\and Mazeh, T.\inst{20} 
\and Moutou, C.\inst{3} 
\and Ofir, A.\inst{20}
\and Ollivier, M.\inst{12} 
\and Parviainen, H. \inst{8,9}
\and P\"atzold, M.\inst{6} 
\and Queloz, D.\inst{10}
\and Rauer, H.\inst{13,21}
\and Rouan, D.\inst{11}
\and Santerne, A. \inst{3}
\and Schneider, J.\inst{22} 
\and Tingley, B.\inst{8,9} 
\and Wuchterl, G.\inst{17} 
}

\institute{
Institut d'Astrophysique de Paris, UMR7095 CNRS, Universit\'e Pierre \& Marie Curie, 98bis Bd Arago, 75014 Paris, France
\and Observatoire de Haute Provence, CNRS/OAMP, 04870 St Michel l'Observatoire, France
\and Laboratoire d'Astrophysique de Marseille, 38 rue Fr\'ed\'eric Joliot-Curie, 13388 Marseille cedex 13, France
\and Universit\'e de Nice-Sophia Antipolis, CNRS UMR 6202, Observatoire de la C\^ote d'Azur, BP 4229, 06304 Nice Cedex 4, France
\and Department of Physics, Denys Wilkinson Building Keble Road, Oxford, OX1 3RH
\and Rheinisches Institut f\"ur Umweltforschung an der Universit\"at zu K\"oln, Aachener Strasse 209, 50931, Germany 
\and McDonald Observatory, The University of Texas, Austin, TX 78712, USA
\and Instituto de Astrofosica de Canarias, E-38205 La Laguna, Tenerife, Spain 
\and Departamento de Astrof\'\i sica, Universidad de La Laguna, E-38200 La Laguna, Tenerife, Spain
\and Observatoire de l'Universit\'e de Gen\`eve, 51 chemin des Maillettes, 1290 Sauverny, Switzerland 
\and LESIA, UMR 8109 CNRS, Observatoire de Paris, UPMC, Universit\'e Paris-Diderot, 5 place J. Janssen, 92195 Meudon, France
\and Institut d'Astrophysique Spatiale, Universit\'e Paris XI, F-91405 Orsay, France 
\and Institute of Planetary Research, German Aerospace Center, Rutherfordstrasse 2, 12489 Berlin, Germany
\and University of Vienna, Institute of Astronomy, T\"urkenschanzstr. 17, A-1180 Vienna, Austria
\and IAG, University of Sao Paulo, Brazil 
\and Research and Scientific Support Department, ESTEC/ESA, PO Box 299, 2200 AG Noordwijk, The Netherlands 
\and Th\"uringer Landessternwarte, Sternwarte 5, Tautenburg 5, D-07778 Tautenburg, Germany
\and University of Li\`ege, All\'ee du 6 ao\^ut 17, Sart Tilman, Li\`ege 1, Belgium
\and Space Research Institute, Austrian Academy of Science, Schmiedlstr. 6, A-8042 Graz, Austria 
\and School of Physics and Astronomy, Raymond and Beverly Sackler Faculty of Exact Sciences, Tel Aviv University, Tel Aviv, Israel  
\and Center for Astronomy and Astrophysics, TU Berlin, Hardenbergstr. 36, 10623 Berlin, Germany
\and LUTH, Observatoire de Paris, CNRS, Universit\'e Paris Diderot; 5 place Jules Janssen, 92195 Meudon, France
}
\date{Received ; accepted }

\abstract{
We report the discovery by the $CoRoT$ space mission of a transiting brown dwarf orbiting 
a F7V star with an orbital period of 3.06 days. CoRoT-15b has a radius of 1.12$^{+0.30}_{-0.15}$ 
{\Rjup}, a mass of 63.3$\pm$4.1 {\Mjup}, and is thus the second transiting companion lying 
in the theoretical mass domain of brown dwarfs. 
CoRoT-15b is either very young or inflated compared to standard evolution models, a situation 
similar to that of M-dwarfs stars orbiting close to solar-type stars. Spectroscopic constraints 
and an analysis of the lightcurve favors a spin period between 2.9 and 3.1 days for the central 
star, compatible with a double-synchronisation of the system.}

\keywords{brown dwarfs - planetary systems - low-mass - techniques: photometry - techniques:
  radial velocities - techniques: spectroscopic}

\titlerunning{CoRoT-15b: a brown dwarf transiting companion}
\authorrunning{}

\maketitle

\section{Introduction}

The $CoRoT$ space mission (Baglin et al. \cite{baglin09}), in operation since the start 
of 2007 and extended to the end of 2013, 
is designed to find transiting exoplanets. A natural product of this mission is that any object 
with size of Jupiter or lower that transits its host star can be detected. This includes 
stellar and sub-stellar companions such as M-dwarfs and brown dwarfs (BDs). 
In the mass-radius diagram {\modif of transiting companions orbiting solar-type stars}, 
there is up until now only one known brown-dwarf, CoRoT-3b 
(Deleuil, et al. 2008), located in the gap in mass between planetary and 
low-mass star companions \footnote{\modif{After submission of our manuscript, Johnson et al. (\cite{johnson10}) reported the discovery of the transiting brown dwarf LHS6343C with a radius of 0.996 $\pm$ 0.026 {\Rjup} and a mass of 70.6 $\pm$ 2.7 \Mjup.}}. Determination of the physical properties of such objects are fundamental 
to understand the link between the population of planets and low-mass stars and 
to distinguish the formation and evolution processes of the two populations. 

We report in this paper the discovery of a new transiting brown-dwarf by $CoRoT$ 
established and characterized thanks to ground-based follow-up observations. 
CoRoT-15b, with an estimated radius of 1.12 {\Rjup} and an estimated mass 
of 63.3 {\Mjup}, orbits in 3.06 days an F-type dwarf with solar metallicity.

\section{$CoRoT$ observations}

SRa02 was the seventh field observed with $CoRoT$ in the second year after its launch. 
It corresponds to the third short run and was located towards the so-called 
galactic anti-center direction. This run started on 2008 October 11 and ended on 2008 
November 12, constituting of a total of 31.7 days of almost-continuous observations.

More than 30 multi-transiting candidates for planets were found among the 10265 targets 
of the SRa02 field. About half of them were clearly identified as binaries 
from light-curve analysis and around tenth of high priority planet-size candidates were 
selected in this short run including SRa02\_E1\_4106 afterwards called CoRoT-15. 
The various ID of this target, including coordinates and magnitudes are listed in 
table~\ref{startable}.

\begin{table}[h]
\caption{ IDs, coordinates and magnitudes.}            
\centering        
\renewcommand{\footnoterule}{}  
\begin{minipage}[!]{7.0cm}  
\begin{tabular}{lc}       
\hline\hline                 
$CoRoT$ window ID &  SRa02\_E1\_4106\\
$CoRoT$ ID &  221686194\\
USNO-B1 ID  & 0961-0097866   \\
2MASS ID   &  06282781+0611105\\
\hline            
RA (J2000)  &  06:28:27.82\\
Dec (J2000) &  +06:11:10.47\\
\hline
B1\footnote{from USNO-B1 catalog}  & 16.85 \\
B2$^a$ & 16.59 \\
R1$^a$  & 15.47 \\
R2$^a$ & 15.43 \\
I$^a$ & 14.83 \\
J\footnote{from 2MASS catalog}  & 13.801 $\pm$ 0.026\\
H$^b$  & 13.423 $\pm$ 0.037 \\
K$^b$  & 13.389 $\pm$ 0.050 \\
\hline
\vspace{-0.6cm}
\end{tabular}
\end{minipage}
\label{startable}      
\end{table}

\section{$CoRoT$ light curve analysis}

\begin{figure}[h]
\centering
\includegraphics[width=8.5cm]{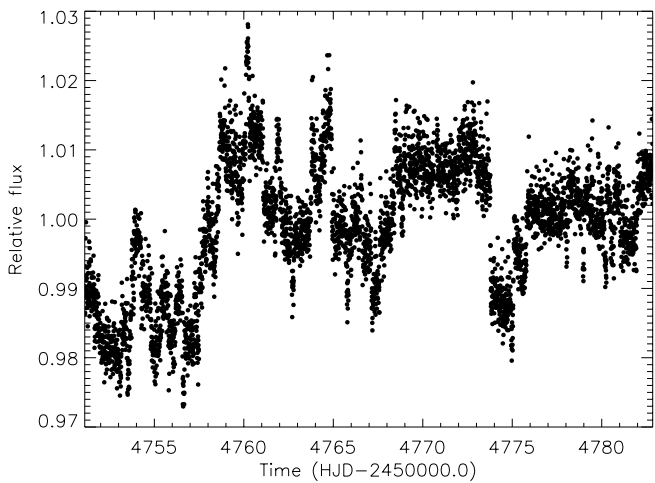}
\caption{Raw light curve of CoRoT-15.}
\label{rawlc}
\end{figure}

CoRoT-15, with an estimated V magnitude close to 16, is located at the faint 
end of the stellar population observed by $CoRoT$. In this magnitude domain, the signal 
is not sufficient to split the photometric aperture in three colors and the extracted 
lightcurve, also called white lightcurve, is monochromatic (Llebaria \& Guterman~\cite{llebaria06}). 
The sampling rate was at 512 seconds during the whole run and not oversampled to 32 
seconds since this candidate was not identified with the alarm mode (Surace et 
al.~\cite{surace08}). Fig.~\ref{rawlc} shows the lightcurve of CoRoT-15 delivered by 
the N2 data levels pipeline. This lightcurve is quite noisy and is affected by several 
high energy particles impacts which result in hot pixels as well as possible stellar variability.

A first analysis of the lightcurve (LC), based on a trapezoidal fit to each individual 
transit, reveals periodic transits with depth of 0.68\% and a period 
of 3.0608$\pm$0.0008 days. 

Since the $CoRoT$ lightcurve is relatively noisy, no meaningful limits to either the 
visible-light albedo of the companion nor to its dayside surface temperature could 
be established.

From the raw lightcurve, we tried to estimate the stellar rotation period. We filtered out 
variations with timescales shorter than 15 data points ($\sim$2.13h) to reduce sensitivity 
to the satellite orbital effects and other short term variations, and we then removed the transits. 
The computed LS periodogram appears to be quite noisy and affected by low-level 
discontinuities in the lightcurve. There is tentative evidence of a possible rotational 
modulation at either 2.9, 3.1 or 6.3 days in the light curve, but the data does not enable 
us to estimate the period more precisely or to distinguish between these values.

\section{Ground-based observations}

\subsection{Ground-based Photometric follow-up}
\label{fuphot}

\begin{figure}[h]
\centering
\includegraphics[width=5cm,angle=-90]{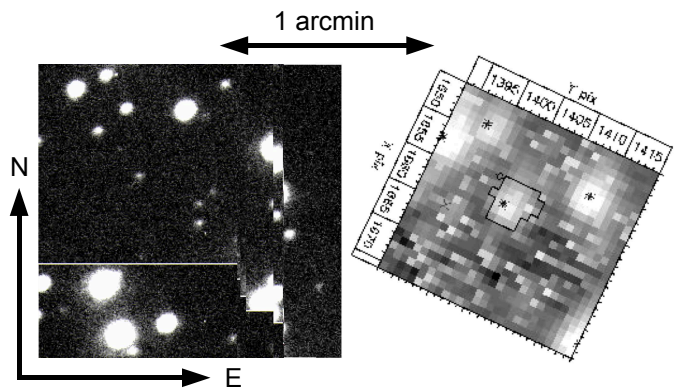}
\caption{The image of the sky around CoRoT-15 (star in the centre). Left:
R-filter image with a resolution of 1.7" taken with the IAC80 telescope. Right:
Image taken by $CoRoT$, at the same scale and orientation. The jagged
outline in its centre is the photometric aperture mask; indicated are
also $CoRoT$'s x and y image coordinates and positions of nearby stars
which are in the Exo-Dat (Deleuil et al 2009) database.}
\label{iac80}
\end{figure}

Ground-based photometry was performed with the aim of refining the target
ephemeris, to verify that none of its closest contaminant stars
correspond to an eclipsing binary and to determine the contamination
from nearby stars inside $CoRoT$'s photometric aperture mask (Deeg et
al.~\cite{deeg09}). CoRoT-15 was observed during a transit event on 2010  
January 13 with time-series photometry at the IAC80 telescope, from HJD 2455209.480 to
.676.  The resultant lightcurve was not sufficiently precise to
identify the expected transit on the target, due to photometric errors
introduced by the presence of thin cirrus. However, the absence of large
brightness variations in the neighboring stars allowed us to exclude
nearby eclipsing binaries as a source of the signals that were
observed by $CoRoT$.
The contamination factor was derived from a measure of the distance and
brightness of the nearby stars on a subset of these IAC80 R-filter
images obtained with the best seeing of that night (1.7 arcsec). Ten nearby stars were 
identified, with six of them contaminating the $CoRoT$ window aperture 
with a flux level that amounts to 1.9$\pm$0.3\% 
of the main target flux. We checked furthermore that none of the known nearby stars 
is bright enough to be contaminating eclipsing binaries. 
The image of the sky around CoRoT-15 is shown in Fig.~\ref{iac80}. 

\subsection{Radial velocity follow-up}

Radial velocity (RV) observations of CoRoT-15 were performed with the HARPS 
spectrograph (Mayor et al. \cite{mayor03}) based on the 3.6-m ESO telescope (Chile) 
as part of the ESO large program 184.C-0639 and with the HIRES spectrograph (Vogt et 
al.~\cite{vogt94}) based on the 10-m Keck-1 telescope as part of NASA's key science 
project to support the $CoRoT$ mission. 

HARPS was used with the observing mode obj\_AB, without simultanenous 
thorium in order to monitor the Moon background light on the second fiber. 
The exposure time was set to 1 hour. A set of 9 spectra was recorded 
between November 24th 2009 and February 21th 2010. 
We reduced HARPS data and computed RVs with the pipeline based on the cross-correlation 
techniques (Baranne et al. \cite{baranne96}; Pepe et al. \cite{pepe02}). 
The signal-to-noise ratio (SNR) per pixel at 550 nm is in the range 3 to 7.8 for this 
faint target. It corresponds to the faint end in magnitude for HARPS follow-up 
observations. Radial velocities were obtained by weighted cross-correlation 
with a numerical G2 mask.  

HIRES observations were performed with the red cross-disperser and the I$_2$-cell to 
measure RVs. We used the 0.861" wide slit that leads 
to a resolving power of $R \approx 45,000$. The contamination of the HIRES spectra 
by scattered moon light was significant for this faint target, but the 7" tall decker 
allowed us to properly correct for the background light. Three spectra without the Iodine cell 
were obtained on 2010 december 2 and january 9. These 3 spectra were co-added to 
serve as stellar template for the RV measurements, and to be used for the determination 
of stellar parameters (see Sect.~\ref{sectspec}).
Over a 4-night run from 2010 January 25 to January 28 we have collected 13 spectra of CoRoT-15 with 
the I$_2$-cell. The average SNR of the spectra with 
the I$_2$-cell range from 13 to 22 (per pixel) in the iodine region from 500-620 nm.
Differential radial velocities were computed using the {\it Austral} Doppler 
code (Endl et al.~\cite{endl00}). Nine RV measurements were made during a transit event 
on 2010 January 25 but were not sensitive enough to detect the signature 
of the Rossiter-McLaughlin effect expected to have, for this high rotating star, 
an amplitude of about 100 {\ms}. 

\begin{table}
\caption{Radial velocity measurements of CoRoT-15 obtained by HARPS and HIRES. 
BJD is the Barycentric Julian Date.}            
\centering                          
\begin{tabular}{lll}       
\hline\hline                 
BJD & RV & $\pm$$1\,\sigma$ \\
-2\,400\,000 & [km\,s$^{-1}$] & [km\,s$^{-1}$]  \\
\hline 
\multicolumn{3}{c}{HARPS 3.6-m ESO} \\
\hline
55159.81312  &   9.944  & 0.308  \\
55169.77922  &   2.107  & 0.315  \\
55235.60498  &   2.849  & 0.380  \\
55236.59903  &   8.057  & 0.392  \\
55241.55686  &   0.119  & 0.338  \\
55243.55544  &   -2.422 & 0.222  \\
55246.56091  &   -2.684 & 0.434  \\
55247.55392  &   -1.318 & 0.411  \\
55248.54372  &   10.235 & 0.357  \\
\hline 
\multicolumn{3}{c}{HIRES 10-m Keck} \\
\hline
55221.76973  &   1.084 &  0.291 \\
55221.81404  &   0.890 &  0.189 \\
55221.82490  &   0.556 &  0.289 \\
55221.89482  &  -0.232 &  0.220 \\
55221.90689  &  -0.492 &  0.216 \\
55221.97788  &  -1.869 &  0.353 \\
55221.99025  &  -1.750 &  0.147 \\
55222.98656  &  -3.193 &  0.340 \\
55222.99752  &  -3.587 &  0.226 \\
55223.73904  &   6.527 &  0.113 \\
55224.82036  &   1.593 &  0.374 \\
55224.83152  &   1.818 &  0.303 \\
55225.04022  &  -1.345 &  0.515 \\
\hline                                   
\end{tabular}
\label{rvtable}      
\end{table}

\begin{figure}[h]
\centering
\includegraphics[width=8.5cm]{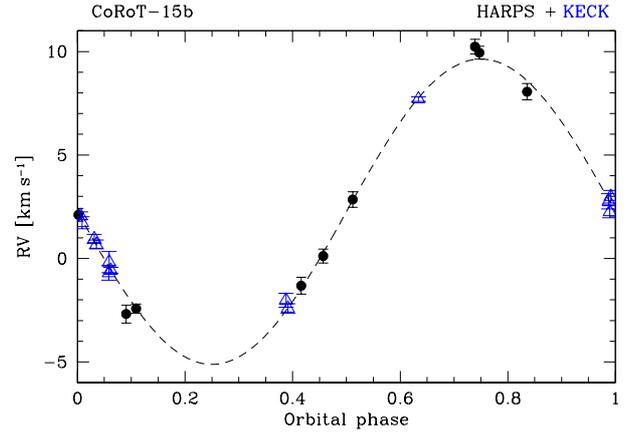}
\caption{Phase-folded radial velocity measurements of CoRoT-15 
with HARPS (dark circle) and HIRES (open triangle)}
\label{rvfig}
\end{figure}

The HARPS and HIRES radial velocities are given in Table~\ref{rvtable}. 
The two sets of relative radial velocities were simultaneously fitted with a Keplerian model,
with the epoch and period of the transit being fixed at the $CoRoT$ value and with an 
adjusted offset between the two different instruments. No significant eccentricity was 
found and we decided to set it to zero. We found a systematic shift in phase using the 
$CoRoT$ period of P=3.0608 days. It comes from the fact that the quite large uncertainty 
on the $CoRoT$ period (69 seconds) may introduce after one year a systematic shift of more 
than 2 hours. When we adjust the period with the RVs fixing the transit epoch 
as determined from $CoRoT$ lightcurve Tt=54753.5570 $\pm$ 0.0028, the best solution is 
obtained for P=3.06039 $\pm$ 0.00014 days and a semi-amplitude $K$=7.376 $\pm$ 0.090 {\kms}. 
The dispersion of the residuals is 0.325 {\kms} and the 
reduced $\chi^2$ is 0.90. The joint analysis of the photometric and RV data, presented in 
Section~\ref{sectparam}, does not change significantly the results.
Figure \ref{rvfig} shows all the radial velocity measurement after subtracting the RV 
offset and phase folded to the updated orbital period.

\subsection{spectral classification}
\label{sectspec}

Three HIRES spectra of CoRoT-15 were acquired without the iodine cell. Each spectrum 
was set in the barycentric rest frame, cleaned from cosmic rays and from the moon reflected 
light. The co-addition of these 3 spectra results in a master spectrum covering the 
wavelength range from 4100 {\AA} to 7800 {\AA} with a SNR per element of resolution in the 
continuum ranging from 20 at 5300 {\AA} up to 70 at 6820 {\AA}. The 9 co-added HARPS spectra 
unfortunately did not permit to reach a better SNR. 

From the analysis of a set of isolated lines, we derived a {\vsini} of 19 $\pm$ 2~\kms.  
The spectroscopic analysis was carried out using the same methodology as for the 
previous $CoRoT$ planets and described in details in Bruntt et al. (\cite{bruntt10}). However, 
the moderate SNR of the master spectrum of this faint target, combined to the marked 
rotational broadening of the spectral lines prevented an accurate measurement of 
the star's photospheric parameters. The derived stellar parameters are reported in 
Table~\ref{starplanet}.

Following Santos et al. (\cite{santos02}) methodology, we also estimated the {\vsini} and 
an $[Fe/H]$ index from the HARPS cross-correlation average parameters (FWHM and surface). 
Assuming a $B-V$ of 0.5, we estimated the {\vsini} of the target to be 16$\pm$1 {\kms}
an $[Fe/H]$ index close to zero (solar metallicity) in agreement with the spectral analysis.

\section{System parameters}
\label{sectparam}

The time span of the CoRoT light curve is relatively short, and the RV data was collected 
on year later. Jointly analysing the two datasets therefore yields significantly improved 
constraints on the period $P$ and time of transit centre $T_{\rm tr}$. To do this, we 
used a Metropolis-Hastings Markov Chain Monte Carlo (MCMC) algorithm (see appendix 1 
of Tegmark et al. \cite{tegmark04} for a general description of MCMC algorithms and 
Winn et al. \cite{winn08} and references therein for a detailed description of their 
application to transits). This has the added advantage of yielding full posterior 
probability distributions for the fitted parameters, ensuring that the effects the well 
known degeneracy between the orbital inclination $i$ and system scale $a/R_\star$ 
(which leads to highly skewed probability distributions for these parameters, as well 
as for the radius ratio $R_{\rm c}/R_\star$) are properly taken into account in the 
final uncertainties.

The light curve was first pre-processed to remove out-of-transit variability as follows. 
Outliers were identified using an iterative non-linear filter (see Aigrain et 
al. \cite{aigrain09}), and a straight line was fitted to the region around each transit. 
{\modif  The effect of contamination reported in section \ref{fuphot}  was taken into account by subtracting a constant amount of flux equal to 1.9\% of the mean flux from the light curve.} 
Each section of the light curve was thus normalised, and a visual check was performed 
to ensure that no residual discontinuities affected the preprocessed light curve. The 
photometric uncertainties were then estimated from the out-of-transit scatter in the 
preprocessed light curve section around each transit. The light curve was modeled using 
the formalism of Mandel \& Agol (\cite{mandel02}). Given the relatively low SNR 
of the transits, we opted to fix the quadratic limb-darkening parameters $u_a$ and 
$u_b$ at the values given by Sing et al. (\cite{sing10}) for the star's effective 
temperature, gravity and metallicity (adopted values: $u_a=0.32$, $u_b=0.30$). The RV 
data were modelled using a Keplerian orbit with eccentricity fixed at zero, since the 
data show no evidence for a significant eccentricity. The relative zero-point of the 
HARPS and HIRES velocities, $\delta V_0$,  was allowed to vary freely. The parameters 
of the MCMC were thus $P$, $T_{\rm tr}$, $R_{\rm c}/R_\star$, $a/R_\star$, the radial 
velocity semi amplitude $K$, the systemic radial velocity $V_{\rm sys}$ and $\delta V_0$.
 
\begin{figure}[h]
\centering
\includegraphics[width=8.5cm]{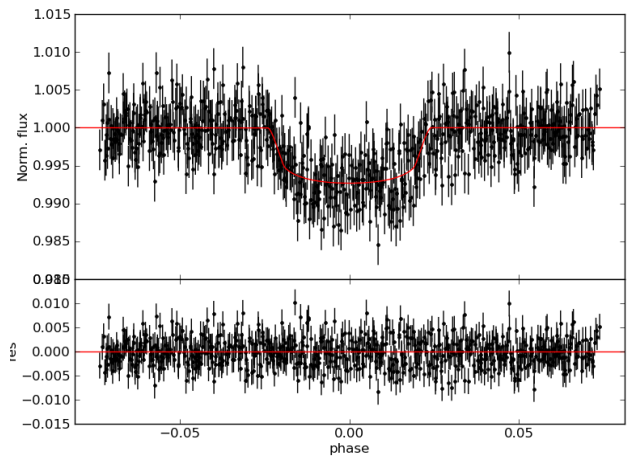}
\caption{Folded, detrended light curve of CoRoT-15 (top), showing the best fit transit 
model (red solid line) and residuals (bottom).}
\label{foldedlc}
\end{figure}

\begin{figure}[h]
\centering
\includegraphics[width=8.5cm]{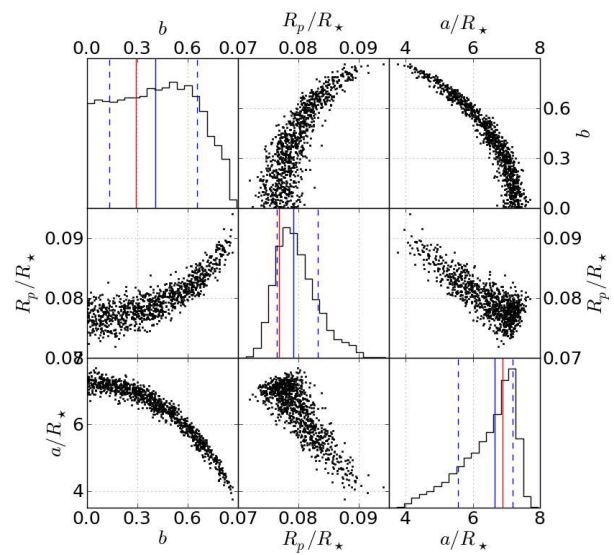}
\caption{Selected posterior probability distributions and two-dimensional correlations for 
the transit fit. The panels along the diagonal show single-parameter posterior probability 
distributions for  $b$, $R_{\rm p}/R_\star$ and $a/R_\star$. The red, blue, and dashed blue 
vertical lines indicate the location of the best-fit, median and the limits of the 68.5\% 
confidence interval for each parameter. The off-diagonal panels show, for each pair of 
parameters, scatter plots of 1000 points randomly selected from the combined MCMC chain. 
The density of the points approximates the joint posterior probability distribution.}
\label{mcmc}
\end{figure}

After an initial chain of $10^5$ steps to adjust the MCMC step sizes for each parameter, 
we ran 10 MCMC chains of $10^5$ steps, each with different starting points. 
The convergence of the chains was checked using the Geldman-Rubin statistic (Geldman 
\& Rubin \cite{gelman92}, Brooks \& Geldman \cite{brooks97}). The chains were 
then combined (after discarding the first $10\%$ of each chain, where the MCMC is 
settling from its starting point) to produce posterior probability distributions for 
each parameters. We report in Table~\ref{starplanet} the median of the probability distribution for 
each parameter\footnote{The choice of which statistic to report is a somewhat tricky one. 
When the distributions are (close to) Gaussian, the median, most probable and 
best-fit values coincide. When the distributions are skewed, as in the case of $b$ for 
example, the median, most probable and best-fit values can differ signficiantly. Whilst 
the best fit value maximises the merit function for the particular dataset being analysed, 
it has no special physical meaning. We adopt the value which divides the probability 
distribution in half, namely the median, as it is arguably the most physically 
meaningful.}. To estimate uncertainties for each parameter, we computed the range of 
values which encloses 68.5\% of the probability distribution (rejecting 16.25\% at each 
extremum). Our uncertainties thus correspond to 68.5\% confidence intervals, just as 
classical 1-$\sigma$ uncertainties do for a Gaussian distribution. The best-fit transit 
model is shown superimposed on the folded light curve in Figure~\ref{foldedlc}. To highlight 
the correlations between $b$, $R_{\rm p}/R_\star$ and $a/R_\star$, and explain the 
rather large resulting uncertainties, we also show the posterior probability 
distributions and 2-$D$ projections of the combined MCMC chain for these parameters 
in Figure~\ref{mcmc}.

We used the photospheric parameters from spectral analysis and the stellar density 
derived from the transit modeling to determine the star's fundamental 
parameters in the (\teff, {\mr}) space. Using {\sl STAREVOL} evolution tracks (Palacios, 
{\sl private com.}), we find the stellar mass to be $M_\star$ = 1.32 $\pm$ 0.10 $M_{\sun}$ 
and the stellar  radius $R_\star$ = 1.46 $^{+0.31}_{-0.14}$ R$_{\sun}$, with an age in the range 
$1.14$--$3.35$~Gyr. This infers a surface 
gravity of \logg\ = 4.23 $^{+0.12}_{-0.20}$, in good agreement with the spectroscopic value.

Calculations using CESAM (Morel \& Lebreton \cite{morel08}, see also Guillot \& Havel 
\cite{guillot10b}) confirm these solutions. The age constraints, 1.9$\pm$1.7~Gyr, are 
however extremely weak, and yield possible pre-main sequence solutions 
with extremely young ages. 

We derived for the transiting companion M$_c$=63.3$\pm$4.1 {\Mjup} and 
R$_c$=1.12$^{+0.30}_{-0.15}$ {\Rjup}.\\

\begin{table}
\centering
\caption{Star and companion parameters.}            
\begin{minipage}[!]{7.0cm} 
\renewcommand{\footnoterule}{}                          
\begin{tabular}{l l}        
\hline\hline                 
Parameters   & Value \\
\hline
Transit epoch $T_{tr}$ [HJD] & 2454753.5608$\pm$0.0011  \\
Orbital period $P$ [days] &  3.06036$\pm$0.00003  \\
Transit duration $d_{tr}$ [h] & 3.24$\pm$0.1 \\
\\
Orbital eccentricity $e$  &  0 (fixed) \\
Semi-amplitude $K$ [\kms] & 7.36$\pm$0.11 \\
Systemic velocity  $V_{0\_harps}$ [\kms] & 2.23$\pm$0.11 \\
Systemic velocity  $V_{0\_hires}$ [\kms] & 1.09$\pm$0.07 \\
\\
Radius ratio $R_{c}/R_{\star}$ & 0.0788$^{+0.0039}_{-0.0029}$ \\
Scaled semi-major axis $a/R_{\star}$ & 6.68$^{+0.49}_{-1.04}$ \\
Impact parameter $b$ & 0.38$^{+0.25}_{-0.26}$\\
\\
{\mr} [solar units]& 0.75$^{+0.05}_{-0.12}$ \\
Stellar density $\rho_{\star}$ [$g\;cm^{-3}$] & 0.60$^{+0.13}_{-0.28}$\\
Inclination $i$ [deg] & 86.7$^{+2.3}_{-3.2}$ \\
\\
Effective temperature $T_{eff}$[K] & 6350$\pm$200 \\
Surface gravity log\,$g$ [dex]&  4.3$\pm$0.2  \\
Metallicity $[\rm{Fe/H}]$ [dex]&  0.1$\pm$0.2 \\
Rotationl velocity {\vsini} [\kms] & 19$\pm$2 \\
Spectral type & F7V\\
\\
Star mass [\Msun] &  1.32$\pm$0.12 \\   
Star radius [\Rsun] &  1.46$^{+0.31}_{-0.14}$  \\
Distance of the system [pc] & 1270$\pm$300 \\
\\
Orbital semi-major axis $a$ [AU] & 0.045$^{+0.014}_{-0.010}$ \\
Companion mass $M_{c}$ [\Mjup] &   63.3$\pm$4.1 \\
Companion radius $R_{c}$[\Rjup]  &  1.12$^{+0.30}_{-0.15}$ \\
Companion density $\rho_{c}$ [$g\;cm^{-3}$] &  59$^{+37}_{-32}$ \\
Equilibrium temperature  $T^{per}_{eq}$ [K] & 1740$^{+120}_{-190}$ \\
\hline       
\vspace{-0.6cm}
\end{tabular}
\end{minipage}
\label{starplanet}     
\end{table}

\section{Discussion and Conclusion}

{\modif CoRoT-15b is one of the rare transiting companion} 
that lies in the theoretical mass domain of brown dwarfs (13-75 MJup, if one adopts the present 
IAU convention). Contrary to CoRoT-3b (Deleuil et al. \cite{deleuil08}) that is located in the overlapping region between the massive planet and the brown-dwarf domain, CoRoT-15b is well in the mass domain of BDs.
Expanding a bit the mass domain, one can easily include in this ensemble the 
high mass "planets" (M $\ge$ 10 {\Mjup}) XO-3b (Johns-Krull et al. \cite{jk08}) and 
WASP-18b (Hellier et al. \cite{hellier09}), and in the M-dwarf regime, OGLE-TR-122b (Pont et 
al. \cite{pont05a}) -123b (Pont et al. \cite{pont06}) -106b (Pont et al. \cite{pont05b}), 
and HAT-TR-205-013 (Beatty et al. \cite{beatty07}). Interestingly, all these objects 
are found to orbit F-type stars (see also Deleuil et al.~\cite{deleuil08}), 
with one exception: OGLE-TR-122b orbits a G-type dwarf but has 
a much longer orbital period (7.3 days compared to less than 4.3 days for all other objects). 

{\modif Early- and mid-F-type} dwarfs have the particularity of being fast rotators, independently of their age 
(Nordstrom et al. \cite{nordstrom97}), a consequence of a small or inexistent outer convective 
zone, weak stellar winds, and reduced losses of angular momentum. 
{\modif The tides raised on a star by its close-in companion (planet, brown dwarf or star) have long been known to pause a threat to its survival (e.g. P\"atzold \& Rauer \cite{patzold02}). This is true when the star's spin is slower than the orbital period of the companion, a common situation for close-in exoplanets. However, massive-enough companions have the possibility of spinning-up the star and may escape engulfment if the total angular momentum of the system is above a critical value (Levrard et al. \cite{levrard09}). Even in that case however, magnetic braking in the central star (e.g. see Barker \& Ogilvie \cite{barker09}) will lead to a loss of angular momentum that will be transferred to the orbit of the companion through tides and lead to orbital decay. We thus propose that close-in massive planets, brown dwarf or M-dwarf can survive when orbiting early or mid F-type dwarfs but that they tend to be engulfed by G-type (or late F-type) dwarfs. In the case of CoRoT-15, we thus expect that the star should be above $\sim$ 1.25 {\Msun}  to avoid efficient spin-down, and that the system should be at or close to double-synchronisation (i.e. the spin period of the star should be close to the orbital period of its companion).}

\begin{figure}[h]
\centering
\includegraphics[width=8.5cm]{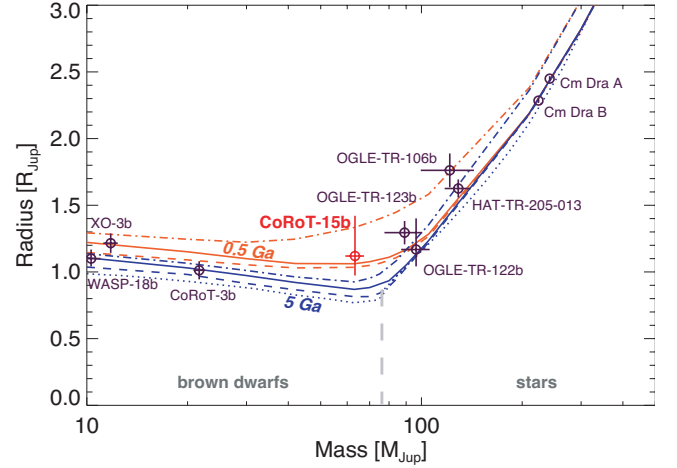}
\caption{
Masses and radii of eclipsing brown-dwarfs and low-mass stars (circles with error bars, as labeled) 
compared to theoretical mass-radius relations (lines). The lines correspond to {\modif isochrones} of 
$0.5$ (upper orange lines) and $5$\,Ga (lower blue lines), respectively. The dashed lines are 
calculated for isolated brown-dwarfs and low-mass stars. The plain lines include the effect of 
irradiation with $T_{\rm eq}=1800\,$K. The dash-dotted lines include irradiation and account for 
a 50\% coverage of the photosphere with low-temperature spots (see text). A 5\,Ga isochrone for 
isolated brown dwarfs/M stars from Baraffe et al. (2003) is shown for comparison (dotted line).}
\label{modelfig}
\end{figure}

It is interesting to see that given the {\vsini} and stellar radius determinations, 
the projected spin period of the central star is $P/sin i_*=3.9^{+0.8}_{-1.1}\,$days. 
An LS periodogram shows the presence of many peaks possibly due to low-level discontinuities 
in the lightcurve. The most robust peak compatible with the $v\sin i$ determination lies 
between $0.32$ and $0.34$ cycles/day, and may thus be linked to a stellar spin period 
between 2.9 and 3.1 days. 
{\modif The CoRoT-15 system thus appears to be indeed close to double-synchronous.} 
Further observations of the system and in particular a precise determination of the stellar spin period 
would be {\modif a powerful mean} of understanding the dynamical evolution of this system. 
{\modif Coupled to studies of similar systems}, this may also yield strong constraints on the 
on the tidal dissipation factor in {\modif F-type dwarfs}.

CoRoT-15b is also extremely interesting for its size in comparison with other objects in this 
mass range, and of evolution tracks for hydrogen-helium brown dwarfs and stars. 
Figure~\ref{modelfig} shows that it {\modif appears} inflated compared to standard evolution tracks 
for these kind of objects (Baraffe et al. \cite{baraffe03}), {\modif although it may be compatible with a young age if the true size is at the lower-end of the one inferred from our measurements. However, we notice that the same problem arises for} OGLE-TR-123b, OGLE-TR-106b and, but to a lesser extent, HAT-TR-205-013. 
\footnote{\modif{The recent discovery of the transiting brown dwarf LHS6343C by Johnson et al. (\cite{johnson10}) 
also points to an inflated companion.}}
{\modif The two other known brown dwarfs with direct radius measurements, discovered in the 
2MASS J05352184-0546085 eclipsing binary system (Stassun et al. \cite{stassun06}), have 
very large radii (5.0 and 6.5 \Rjup) but related to the very young age of the system ($\sim$ 1 Myr), still 
in the earliest  stages of gravitational contraction.}

In order to examine possible solutions 
to this puzzle {\modif (other than a systematic overestimation of the inferred sizes for the systems known thus far)}, 
we calculate evolution tracks using CEPAM (Guillot \& Morel \cite{guillot95}), 
but adding the dominant thermonuclear reaction cycle, namely the pp-chain (see Burrows \& Liebert 
\cite{burrows93}). The atmospheric boundary condition is adjusted to the Baraffe et al. 
(\cite{baraffe03}) evolution tracks, using the analytical solution of Guillot (\cite{guillot10a}), 
and values of the thermal and visible mean opacities, $\kappa_{\rm th}=0.04\,\rm cm^2\,g^{-1}$ 
and $\kappa{\rm v}=0.024\,\rm cm^2\,g^{-1}$. The model shows that irradiation effects, although 
quite significant for Jupiter-like planets have rather small consequences in the brown dwarf regime, 
and become completely negligible in the stellar regime. We also test the possibility that these 
inflated sizes may be explained by the presence of cold spots on the brown dwarf, similarly to a 
mechanism proposed to explain that M-type star in close-in binaries also appear inflated 
(Chabrier et al. \cite{chabrier07}). As shown in Fig.~\ref{modelfig}, this mechanism works only 
in combination with a young age for the system, and with a rather large fraction ($\sim$50\%) of the 
photosphere covered with spots. An alternative possibility is that irradiated atmospheres are much 
more opaque than usually thought, possibly a consequence of photochemistry and disequilibrium 
chemistry. {\modif A solar metallicity was assumed for the brown-dwarf models displayed in Fig. 6. 
We tested the effect of metallicity on the mean molecular mass and the opacity but did not find 
significant change in the radius.}

In any case, this shows that CoRoT-15b is a crucial object to understand both the dynamical 
and physical evolution of giant planets, brown-dwarfs and low-mass stars. Further observations 
aiming at obtaining more accurate spectra of the star would be highly desirable. Although this 
is a challenging measurement given the faintness of the target, measurement of secondary transits 
in the infrared would be extremely interesting because they would inform us on this rare heavily 
irradiated brown dwarf atmosphere.

\begin{acknowledgements}
The authors wish to thank the staff at ESO La Silla Observatory for their support 
and for their contribution to the success of the HARPS project and operation. 
HIRES data presented herein were obtained at the W.M. Keck Observatory from telescope time 
allocated to the NASA through the agency's scientific partnership with the California 
Institute of Technology and the University of California. 
The French team wish to thank the Programme National de Plan\'etologie (PNP) 
of CNRS/INSU and the French National Research Agency (ANR-08-JCJC-0102-01) for 
their continuous support to our planet search program. 
The team at IAC acknowledges support by grant ESP2007- 65480-C02-02 of
the Spanish Ministerio de Ciencia e Innovaci\'on.
The German CoRoT Team (TLS and the University of Cologne)
acknowledges DLR grants 50OW0204, 50OW0603, and 50QP07011.
MG acknowledges support from the Belgian Science Policy
Office in the form of a Return Grant.
SA \& NG acknowledge support from STFC standard grand ST/G002266.
FB acknowledges the continuous support of PLS-230371.

\end{acknowledgements}

\end{document}